\documentclass[12pt]{article}
\usepackage{latexsym}

\newcommand{\lbl}[1]{\label{eq:#1}}
\newcommand{ \rf}[1]{(\ref{eq:#1})}

\newcommand{\be}{\begin{equation}}
\newcommand{\ee}{\end{equation}}
\newcommand{\bea}{\begin{eqnarray}}
\newcommand{\eea}{\end{eqnarray}}
\newcommand{\noi}{\noindent}
\newcommand{\nn}{\nonumber}
\newcommand{\ra}{\rightarrow}

\newcommand{\lesssim}{ {\
\lower-1.2pt\vbox{\hbox{\rlap{$<$}\lower5pt\vbox{\hbox{$\sim$}}}}\ } 
}
\newcommand{\gtrsim}{ {\
\lower-1.2pt\vbox{\hbox{\rlap{$>$}\lower5pt\vbox{\hbox{$\sim$}}}}\ } 
}

\newcommand{\cL}{{\cal L}}

\newcommand{\cO}{{\cal O}}

\newcommand{\tr}{\mbox{\rm tr}}

\newcommand{\MeV}{\mbox{\rm MeV}}
\newcommand{\GeV}{\mbox{\rm GeV}}

\newcommand{\annd}{\mbox{\rm and}}

\newcommand{\eff}{\mbox{\rm {\scriptsize eff}}}
\newcommand{\exxp}{\mbox{\rm exp.}}

\input epsf

\title{Large--$N_c$ QCD and Weak Matrix Elements}

\author{Eduardo~de Rafael\thanks{CPT,
CNRS--Luminy, Case 907 F-13288 Marseille Cedex 9, France}}

\begin{document}

\maketitle

\begin{abstract}
I report on recent progress~\cite{eder:KPdeR98,eder:KPPdeR99} in
calculating electroweak processes within the framework of QCD in the
$1/N_c$ expansion. 
\end{abstract}

\vspace*{0.3cm}
\noi
{\bf 1. Introduction}

\noi
In the Standard Model, the physics of non--leptonic $K$--decays is
described by an
effective Lagrangian which is the sum of four--quark operators
modulated by $c$--number coefficients (Wilson coefficients).
This effective Lagrangian results
from integrating out the fields in the Standard Model with heavy masses
$(Z^0,W^{\pm},t,b$ and $c$), in the presence of the strong
interactions evaluated in
perturbative QCD (pQCD) down to a scale
$\mu$ below the mass of the charm quark $M_{c}$. The scale $\mu$ has 
to be large enough for the pQCD evaluation of the $c$--number coefficients
to be valid and, therefore, it is much larger  than the scale at which an
effective Lagrangian description in terms of the Nambu--Goldstone degrees
of freedom ($K$,
$\pi$ and $\eta$) of the  spontaneous $SU(3)_{L}\times SU(3)_{R}$ 
symmetry breaking (S$\chi$SB) is appropriate. Furthermore, the
evaluation of the coupling constants of the low--energy effective
chiral Lagrangian cannot be made within pQCD
because at scales
$\mu\lesssim 1\,\GeV$ we enter a regime where S$\chi$SB and confinement
take place
and the dynamics of QCD is then fully governed by non--perturbative 
phenomena.

The structure of the low--energy effective Lagrangian, in the absence of
virtual electroweak interactions, is well--known~\cite{eder:Wei67}
\be\lbl{GL}
\cL_{\eff}=\frac{1}{4}f_{\pi}^2\ \tr D_{\mu}UD^{\mu}U+\cdots+
L_{10}\ \tr U^{\dagger}F_{R}^{\mu\nu}UF_{L\mu\nu}+\cdots\,.
\ee
Here the unitary matrix $U$ collects the meson fields ($K$,
$\pi$ and $\eta$) and $F_{L}$, $(F_{R})$ denote
field--strength tensors associated to external gauge field sources. The
dots indicate other terms with the same chiral power counting
$\cO (p^4)$ as the $L_{10}$ term and
higher order terms. The important point that I wish to emphasize here is
that {\it the coupling constants of this effective Lagrangian, correspond to
coefficients of the Taylor expansion in powers of momenta (and quark
masses), of specific QCD Green's functions of colour singlet
quark--currents.} Let us consider as an example, and in the chiral limit
where the light quark masses are set to zero, the two--point function ($Q^2=
-q^2$; $L^{\mu}=\bar{q}\gamma^{\mu}\frac{1}{2}(1-\gamma_{5})q$; 
$R^{\mu}=\bar{q}\gamma^{\mu}\frac{1}{2}(1+\gamma_{5})q$)
\be\lbl{lrtpf}
\Pi_{LR}^{\mu\nu}(q)=2i\!\int d^4 x\,e^{iq\cdot x}\langle 0\vert
\mbox{\rm T}\!\left(L^{\mu}(x)R^{\nu}(0)^{\dagger}
\right)\!\vert 0\rangle=(\!q^{\mu}q^{\nu}\!-\!g^{\mu\nu}q^2)\Pi_{LR}(Q^2)\,.
\ee
For $Q^2$ small, $-Q^2\Pi_{LR}(Q^2)=f_{\pi}^2 + 4L_{10}\ Q^2 +\cO(Q^4)$, 
clearly showing the correspondence stated above.

In the presence of virtual electroweak interactions there appear new
couplings in the low--energy effective Lagrangian, like e.g. the term 
\be\lbl{effem}
e^2 C\,\tr
\left(Q_{R}UQ_{L}U^{\dagger}\right)=
-2e^2 C\frac{1}{f_{\pi}^2} (\pi^+
\pi^- + K^+ K^-) + \cdots\,, 
\ee
where $Q_{R}\!=\!Q_{L}\!=\!\mbox{\rm diag}[2/3, -1/3, -1/3]$, 
showing that, in the presence of the electroweak interactions,
the charged pion and kaon fields become massive. The basic complication 
in evaluating coupling constants like $C$ in Eq.~\rf{effem}, which originate
in loops with electroweak gauge fields,  is
that {\it they correspond to integrals over \underline {all values} of the
euclidean momenta of specific combinations of QCD Green's functions of
colour singlet quark--currents}. In our particular example, it can be shown
~\cite{eder:Lowetal67,eder:KPdeR98} that
\be\lbl{piew}
C=\frac{-1}{8\pi^2}\,\frac{3}{4}\,
\int_0^\infty dQ^2\,Q^2\left(1-\frac{1}{Q^2
+M_{Z}^2}\right)\Pi_{LR}(Q^2)\,,
\ee
with $Q$ the euclidean momentum of the virtual gauge field;
the first term in the parenthesis is the well known~\cite{eder:Lowetal67}
contribution from electromagnetism; the second term is the one induced
by the weak neutral current~\cite{eder:KPdeR98}. It is clear that the
evaluation of coupling constants of this type
represents a rather formidable task. As we shall see below, it is
possible, however, to proceed further  within the framework of the
$1/N_{c}$--expansion in QCD~\cite{eder:tH74}.  

\vspace*{0.3cm}
\noi
{\bf 2. Large--$N_c$ QCD and the OPE}

\noi
In the limit where the
number of colours
$N_c$ becomes infinite, with $\alpha_{s}\times N_{c}$ fixed,
the QCD spectrum reduces to an infinite number of zero--width 
mesonic resonances, and the leading large--$N_c$
contribution to an $n$--point correlator is given by all the possible
tree--level  exchanges of these resonances in the various channels. 
In this limit, the
analytical structure of an $n$--point function is very
simple:~the  singularities in each channel consist only of a succession
of {\it simple poles}. For example, in the case of $\Pi_{LR}$ in
Eq.~\rf{lrtpf},
\be\lbl{LRN1}
-Q^2\Pi_{LR}(Q^2)=f_{\pi}^2+\sum_{A}f_{A}^2
M_{A}^2\frac{Q^2}{M_{A}^2+Q^2} -\sum_{V}f_{V}^2
M_{V}^2\frac{Q^2}{M_{V}^2+Q^2}\,,
\ee
where the sums extend over all vector ($V$) and axial--vector ($A$)
states. Furthermore, in the chiral limit, the operator product
expansion (OPE) applied to the correlation function $\Pi_{LR}(Q^2)$
implies
\be
\lim_{Q^2\ra\infty} Q^2\Pi_{LR}(Q^2)\ra 0\,,\quad
\lim_{Q^2\ra\infty} Q^4\Pi_{LR}(Q^2)\ra 0\,, 
\ee
and~\cite{eder:SVZ79}
\be\lbl{d6}
\lim_{Q^2\ra\infty}Q^6\Pi_{LR}(Q^2)=-4\pi^2\left(\frac{\alpha_s}
{\pi}+
\cO(\alpha_s^2)\right)
\langle\bar{\psi}\psi\rangle^2\,.
\ee
The first two relations result in the two Weinberg sum
rules
\be\lbl{weinbergsrs}
\sum_{V}f_{V}^2 M_{V}^2-\sum_{A}f_{A}^2 M_{A}^2=f_{\pi}^2
\quad
\annd\quad
\sum_{V}f_{V}^2 M_{V}^4-\sum_{A}f_{A}^2 M_{A}^4=0\,.
\ee
There is in fact a new set of
constraints that emerge in the large--$N_c$ limit which relate order
parameters of the OPE to couplings and masses of the narrow
states. In our example, we have from Eqs.~\rf{LRN1} and~\rf{d6}, that 
\be
\sum_{V} f_{V}^2
M_{V}^6-\sum_{A} f_{A}^2 M_{A}^6
=-4\pi^2\left(\frac{\alpha_s}
{\pi}+
\cO(\alpha_s^2)\right)
\langle\bar{\psi}\psi\rangle^2\,.
\ee
On the other hand, the coupling constants of the low--energy
Lagrangian in the strong interaction sector are also related to couplings
and masses of the narrow states of the large--$N_c$ QCD spectrum; e.g.,
\be\lbl{L10}
-4L_{10}=\sum_{V} f_{V}^2-\sum_{A}f_{A}^2\,.
\ee 

It is to be remarked that the convergence of the integral in Eq.~\rf{piew}
in the large--$N_c$ limit is guaranteed by the two Weinberg sum rules in
Eqs.~\rf{weinbergsrs}. However, in order to obtain a numerical estimate,
and in the absence of an explicit solution of
QCD in the large--$N_c$ limit, one still needs to make further
approximations. Partly inspired by the
phenomenological successes of ``vector meson dominance'' in predicting
e.g., the low--energy constants of the effective chiral
Lagrangian~\cite{eder:EGLPR89}, we have recently
proposed~\cite{eder:PPdeR98} to consider the approximation to
large--$N_c$ QCD, which restricts the hadronic spectrum
to a minimal pattern, compatible with the short--distance 
properties of the QCD Green's functions which govern the observable(s) one
is interested in. In the channels with $J^P$ quantum numbers $1^{-}$ and
$1^{+}$ this minimal pattern, in the cases which we have discussed so far,
is the one with a spectrum which consists of a hadronic lowest energy
narrow state and treats the rest of the narrow states as a large--$N_c$ pQCD
continuum; the onset of the continuum being fixed by consistency
constraints from the OPE, like the absence of dimension
$d=2$ operators. We call this the {\it lowest meson dominance} (LMD)
approximation to large--$N_c$ QCD. The basic observation here is that {\it
order parameters of S$\chi$SB in QCD have a smooth behaviour at short
distances}. For example, in the case of the function
$\Pi_{LR}$ and, therefore, the coupling $C$, this is reflected by the fact
that (in the chiral limit) the pQCD continuum contributions in the $V$--sum
and the $A$--sum in Eq.~\rf{LRN1} cancel each other. The evaluation of the
constant $C$ in Eq.~\rf{piew} in this approximation, corresponds to a
mass difference  $\Delta m_{\pi}=4.9\,\MeV$, remarkably close to the
experimental result: $\Delta m_{\pi}\vert_{\exxp}=4.59\,\MeV$.

\vspace*{0.3cm}
\noi
{\bf 3. Electroweak Penguin Operators.}

\noi
Within the framework discussed above, we have also
shown~\cite{eder:KPdeR98} that the $K\to\pi\pi$ matrix elements of the
four--quark operator
\be\lbl{Q7}
Q_7 = 6(\bar{s}_{L}\gamma^{\mu}d_{L})
\sum_{q=u,d,s} e_{q} (\bar{q}_{R}\gamma_{\mu}q_{R})\,,
\ee
generated by the electroweak penguin--like diagrams of
the Standard Model, can be calculated to first non--trivial order in the
chiral expansion and in the $1/N_c$ expansion. 
What is needed here is the bosonization of the operator $Q_{7}$ to
next--to--leading order in the $1/N_c$ expansion. The problem
turns out to be entirely analogous to the bosonization of the operator
$Q_{LR}\equiv\left(\bar{q}_{L}\gamma^{\mu}Q_{L}q_{L}\right)
\left(\bar{q}_{R}\gamma^{\mu}Q_{R}q_{R}\right)$ which governs the 
electroweak $\pi^{+}\!-\!\pi^{0}$ mass difference  discussed
above. Because of the $LR$ structure, the factorized  component of $Q_{7}$,
which is leading in $1/N_c$, cannot contribute to order
$\cO(p^0)$ in the low--energy effective Lagrangian. The first $\cO(p^0)$
contribution from this operator is next--to--leading in the $1/N_c$
expansion and is given by an integral,
[$\left(\lambda_{L}^{(23)}\right)_{ij}=\delta_{i2}\delta_{3j}$
$(i,j=1,2,3)$,] 
\be\lbl{LRgral}
Q_{7}\ra -3ig_{\mu\nu}\int \frac{d^4q}{(2\pi)^4}
\Pi_{LR}^{\mu\nu}(q)\,\,
\tr\left( U\lambda_{L}^{(23)} U^{\dagger} Q_{R}\right)\,,
\ee
involving the \underline {same} two--point function as in Eq.~\rf{lrtpf}.
Although the resulting $B$ factors of $\Delta I\!=\!1/2$ and
$\Delta I\!=\!3/2$ transitions  are found to depend only logarithmically on
the matching scale $\mu$, their actual numerical values turn out to be 
rather sensitive to the precise choice of $\mu$ in the $\GeV$ region.
Furthermore, because of the normalization to the vacuum saturation
approximation (VSA) inherent to the (rather disgraceful) conventional
definition of $B$--factors, there appears a spurious dependence on the
light quark masses as well. In Fig.~1 we show our prediction for the ratio
\be\lbl{lattice}
{\widetilde B}_{7}^{(3/2)}\,\equiv\,\frac{\langle\pi^+
\vert Q_7^{(3/2)}\vert K^+\rangle}{\langle\pi^+\vert Q_7^{(3/2)}\vert
K^+\rangle^{\rm VSA}_0}\,,
\ee
versus the matching scale $\mu$ defined in the ${\overline{MS}}$ scheme.
This is the ratio considered in recent lattice QCD
calculations~\cite{eder:lattice}. [In fact, the lattice definition of
${\widetilde B}_{7}^{(3/2)}$ uses a current algebra relation between the
$K\to\pi\pi$ and the $K\to\pi$ matrix elements which is only valid at
order $\cO (p^0)$ in the chiral expansion.]
In Eq.~\rf{lattice}, the matrix element in the denominator is evaluated 
in the chiral limit, as indicated by the subscript ``0''.

\vspace{0.5cm}
\centerline{\epsfbox{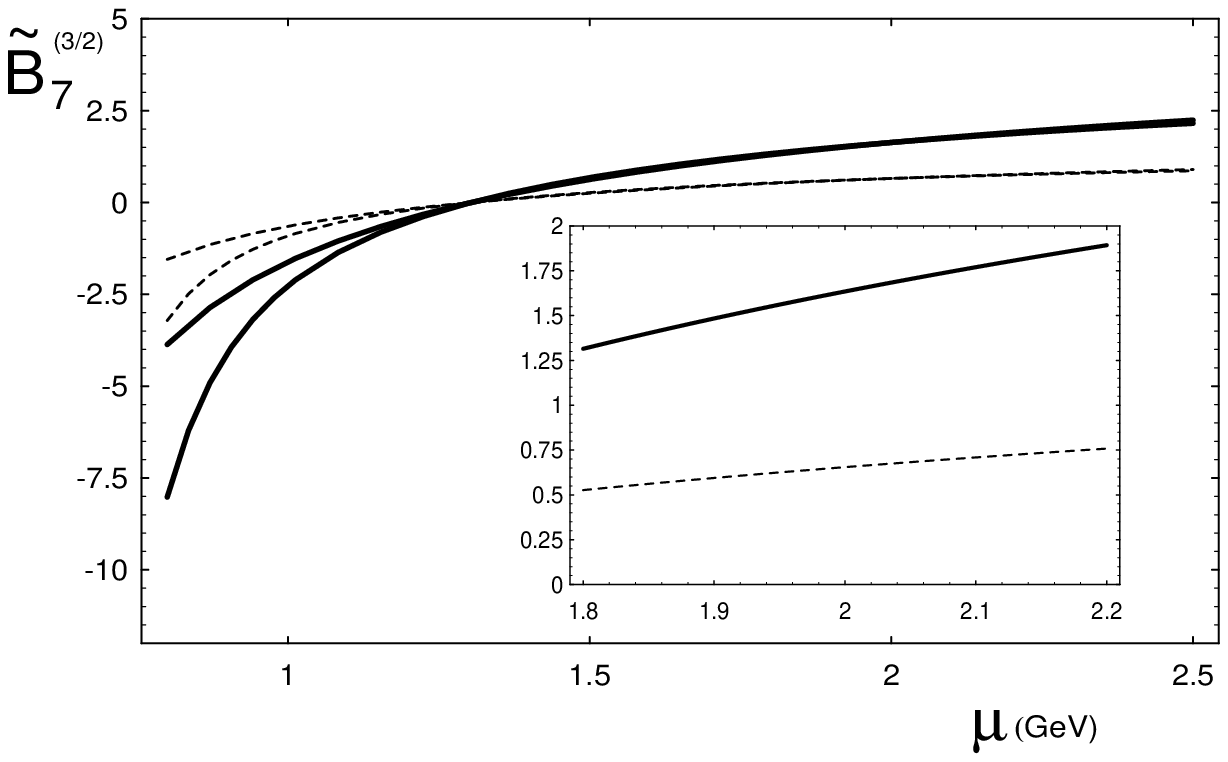}}
\noi
{ Fig.~1:} {\small The $\tilde{B}_{7}^{(3/2)}$ factor in
Eq.~\rf{lattice} versus $\mu$ in $\GeV$. Solid lines correspond to
$(m_s \!+ \!m_d)(2\,{\rm GeV}) = 158\,{\rm MeV}$; dashed lines
to $(m_s + m_d)(2\,{\rm GeV}) = 100\,{\rm MeV}$.}

\vspace*{0.3cm}
\noi
{\bf 4. Decay of Pseudoscalars into Lepton Pairs.}

\noi
The processes $\pi\to e^+ e^-$ and $\eta\to l^+ l^-$ ($l=e,\mu$) are
dominated by the exchange  of two virtual photons. It is then
useful to  consider the ratios ($P=\pi^0,\eta$)
\be\lbl{ratio}
R(P\to\ell^+\ell^-) = \frac{Br (P\to\ell^+\ell^-)}{Br (P\to\gamma\gamma)}
=
2\bigg(\frac{\alpha m_{\ell}}{\pi M_P}\bigg)^2\,
\beta_{\ell}(M_P^2)\,\vert{\cal A}(M_P^2)\vert^2,\lbl{width}
\ee
with $\beta_{\ell}(s) = \sqrt{1-4m_{\ell}^2/s}$.
To lowest order in the
chiral expansion, the unknown dynamics in the amplitude ${\cal A}(M_P^2)$
depends entirely on a low--energy coupling constant 
$\chi$. We have recently shown~\cite{eder:KPPdeR99} that this constant can
be expressed as an integral over the three--point function
\bea\lbl{VVP}
\lefteqn{\!\!\!\!\!\!\!\!\!\!\!\!\!\!\!\!\!\!\int d^4x\int
d^4y\,e^{iq_1\cdot x}e^{iq_2\cdot y}
<0\,\vert\,T\{j_\mu^{\mbox{\scriptsize{em}}}(x)
              j_\nu^{\mbox{\scriptsize{em}}}(y)P^3(0)\}\,\vert\,0>}\nn \\
 & & 
\ \ \ \ \ \ \  =\frac{2}{3}\,\epsilon_{\mu\nu\alpha\beta}q_1^\alpha
q_2^\beta
\,{\cal H}(q_1^2,q_2^2,(q_1+q_2)^2)\,,
\eea
involving the electromagnetic current $j_\mu^{\mbox{\scriptsize{em}}}$
and the density current 
$P^3\,\!=\!{1\over 2}({\bar u}i\gamma_5 u - {\bar d}i\gamma_5 d)$.
More precisely, ($d\!=\!$ space--time dimension,)
\bea\lbl{chisimp}
\lefteqn{\frac{\chi(\mu)}{32\pi^4}
\frac{<{\bar \psi}\psi>}{F_{\pi}^2}
=\ -\bigg(1\,-\,\frac{1}{d}\bigg)\,\int \frac{d^dq}{(2\pi)^d}\,
\bigg(\frac{1}{q^2}\bigg)^2\times} \nn \\
 & & 
\lim_{(p'- p)^2\to 0}\,
(p'-p)^2\bigg[\,{\cal H}(q^2,q^2,(p'-p)^2)
-{\cal H}(0,0,(p'-p)^2)\,\bigg]\,.
\eea
The evaluation of this coupling in the LMD approximation to large--$N_c$ QCD
which we have discussed above, leads to the result
$\chi^{\mbox{\scriptsize LMD}}(\mu \!= \!M_V)\ =\ 2.2\pm 0.9$.
The corresponding branching ratios are shown in Table~1.
\begin{table}[!h]
\caption{\small Ratios $R(P\to\ell^+\ell^-)$ in Eq.~\rf{ratio} obtained 
within the LMD approximation to large--$N_C$ QCD and the comparison with 
available experimental results.}
\begin{center}
\begin{tabular}{ccc}
\hline\hline
$R$ & LMD & Experiment\\ \hline
$R(\pi^0\to e^+e^-)\times 10^{8}$ & $6.2\pm 0.3$ & 
$7.13\pm 0.55$~\cite{eder:AH99}
  \\ \hline
$R(\eta\to \mu^+\mu^-)\times 10^{5}$ & $1.4\pm 0.2$ & $1.48\pm 0.22
$~\cite{eder:PDG98} \\ \hline
$R(\eta\to e^+e^-)\times 10^{8}$ & $1.15\pm 0.05$ & ?\\
\hline\hline 
\end{tabular}
\end{center}
\label{table1}
\end{table}

It was shown in ref.~\cite{eder:GDP98} that, when evaluated within
the  chiral $U(3)$ framework and in the $1/N_c$ expansion, the 
$\vert\Delta S\vert = 1$ $K_L^0\to\ell^+\ell^-$ 
transitions can also be described by an expression like in Eq.~\rf{ratio}
with an effective constant 
$\chi_{K^0_L}$ containing an additional piece
from the short--distance contributions~\cite{eder:BF97}.
The most accurate experimental determination~\cite{eder:Amb99} gives: 
$Br(K_L^0\to\mu^+\mu^-)=(7.18\pm 0.17)\times 10^{-9}$. 
In the framework of the
$1/N_c$ expansion and using the experimental branching
ratio~\cite{eder:PDG98} 
$Br(K_L^0\to\gamma\gamma)=(5.92\pm 0.15)\times 10^{-4}$, this 
leads to a unique solution for an
{\it effective} $\chi_{K_L^0}=5.17\pm 1.13$. Furthermore, 
following Ref.~\cite{eder:GDP98},  
$\chi_{K_L^0}=\chi - {\cal N}\ \delta \chi_{SD}$ where 
${\cal N}=(3.6/g_8 c_{\rm red})$  normalizes the 
$K_L^0\to\gamma\gamma$ amplitude. The coupling $g_8$ governs 
the $\Delta I\!=\!1/2$
rule, the constant $c_{\rm red}$ is defined in Ref.~\cite{eder:GDP98} and  
$\delta \chi_{SD}^{\mbox{\tiny\rm Standard}}=(+1.8\pm 0.6)$ is the
short--distance contribution in the Standard Model~\cite{eder:BF97}. 
Therefore, a test of the {\it short--distance}   contribution to
this process completely hinges on our understanding of the {\it
long--distance} constant ${\cal N}$ and therefore of the $\Delta I\!=\!1/2$
rule in the $1/N_c$ expansion. Moreover, $c_{\rm red}$ is regrettably  very
unstable in the chiral and large--$N_c$ limits, a behaviour that surely
points  towards the need to have higher order corrections under control.
The analysis of Ref.~\cite{eder:GDP98}  uses $c_{\rm red}\simeq +1$ and
$g_8\simeq 3.6$, where these numbers are obtained phenomenologically by
requiring agreement  with  the two--photon decay of $K_L^0, \pi^0, \eta$
and $\eta'$ as well  as $K\to 2\pi, 3\pi$. Should we use these values of
$c_{\rm red}$ and $g_8$  with our result  $\chi^{\mbox{\scriptsize
LMD}}(\mu \!= \!M_V)\ =\ 2.2\pm 0.9$ we would obtain $\chi_{K_L^0}=0.4\pm
1.1$,  corresponding to a ratio 
$R(K_L^0\to\mu^+ \mu^-)=(2.24\pm 0.41)\times 10^{-5}$ which is $2.5\sigma$ 
above the experimental value 
$R(K_L^0\to\mu^+ \mu^-)=(1.21\pm 0.04)\times 10^{-5}$.  

\vspace*{0.3cm}
\noi
{\bf Acknowledgements}

\noi
It is a pleasure to thank my colleagues Marc Knecht, Santi Peris and
Michel Perrottet
for a very pleasant collaboration. This research
was supported, in part, by TMR, EC--Contract No. ERBFMRX--CT980169.

\end{document}